\documentclass[aps,prc,reprint,superscriptaddress,showpacs,nofootinbib]{revtex4-1}
\usepackage{amsmath,amsfonts,amssymb,amscd,amsthm}
\usepackage[utf8]{inputenc}
\usepackage{epsfig}
\usepackage{graphicx}
\usepackage{xcolor}

\usepackage{hyperref}
\hypersetup{backref=true,       
    pagebackref=true,               
    hyperindex=true,                
    colorlinks=true,                
    breaklinks=true,                
    urlcolor= blue,  
    linkcolor= blue, 
    bookmarks=blue,                 
    bookmarksopen=false,
    filecolor=black,
    citecolor=blue, 
    linkbordercolor=blue
}

\usepackage{array}
\usepackage[ulem=normalem]{changes}

\newcommand{\Ca}{$^{36}$Ca\,}

\begin{document}

\title{Evaluation of the  $^{35}$K($p$,$\gamma$)$^{36}$Ca reaction rate using the  $^{37}$Ca($p$,$d$)$^{36}$Ca transfer reaction}

\author{L.~Lalanne}
    \email[]{louis.lalanne@ijclab.in2p3.fr}
    \affiliation{Universit\'e Paris-Saclay, CNRS/IN2P3, IJCLab, 91405 Orsay, France}
    \affiliation{Grand Accélérateur National d'Ions Lourds (GANIL), CEA/DRF-CNRS/IN2P3, 
    Bd. Henri Becquerel, 14076 Caen, France}  
\author{O.~Sorlin}
    \email[]{olivier.sorlin@ganil.fr}
    \affiliation{Grand Accélérateur National d'Ions Lourds (GANIL), CEA/DRF-CNRS/IN2P3, 
    Bd. Henri Becquerel, 14076 Caen, France}
\author{M.~Assi\'e}
    \affiliation{Universit\'e Paris-Saclay, CNRS/IN2P3, IJCLab, 91405 Orsay, France}
\author{F.~Hammache}
    \affiliation{Universit\'e Paris-Saclay, CNRS/IN2P3, IJCLab, 91405 Orsay, France}
\author{N.~de~S\'{e}r\'{e}ville}
    \affiliation{Universit\'e Paris-Saclay, CNRS/IN2P3, IJCLab, 91405 Orsay, France}
\author{S.~Koyama}
    \affiliation{Department of Physics, Unviversity of Tokyo}
     \affiliation{Grand Accélérateur National d'Ions Lourds (GANIL), CEA/DRF-CNRS/IN2P3, 
    Bd. Henri Becquerel, 14076 Caen, France}
\author{D.~Suzuki}
    \affiliation{RIKEN Nishina Center, 2-1, Hirosawa, Wako, Saitama 351-0198, Japan}
\author{F.~Flavigny}
    \affiliation{Universit\'e Paris-Saclay, CNRS/IN2P3, IJCLab, 91405 Orsay, France}
    \affiliation{LPC Caen, Normandie Université, ENSICAEN, UNICAEN, CNRS/IN2P3, Caen, France}
\author{D.~Beaumel}
    \affiliation{Universit\'e Paris-Saclay, CNRS/IN2P3, IJCLab, 91405 Orsay, France}
\author{Y~Blumenfeld}
    \affiliation{Universit\'e Paris-Saclay, CNRS/IN2P3, IJCLab, 91405 Orsay, France}
\author{B.~A.~Brown}
    \affiliation{Department of Physics and Astronomy, National Superconducting Cyclotron Laboratory,
Michigan State University, East Lansing, Michigan}
\author{F.~De Oliveira Santos}
 \affiliation{Grand Accélérateur National d'Ions Lourds (GANIL), CEA/DRF-CNRS/IN2P3,
    Bd. Henri Becquerel, 14076 Caen, France}
\author{F.~Delaunay}
    \affiliation{LPC Caen, Normandie Université, ENSICAEN, UNICAEN, CNRS/IN2P3, Caen, France}
\author{S.~Franchoo}
    \affiliation{Universit\'e Paris-Saclay, CNRS/IN2P3, IJCLab, 91405 Orsay, France}
\author{J.~Gibelin}
    \affiliation{LPC Caen, Normandie Université, ENSICAEN, UNICAEN, CNRS/IN2P3, Caen, France}
\author{V.~Girard-Alcindor}
    \affiliation{Grand Accélérateur National d'Ions Lourds (GANIL), CEA/DRF-CNRS/IN2P3, 
    Bd. Henri Becquerel, 14076 Caen, France}
\author{J.~Guillot}
    \affiliation{Universit\'e Paris-Saclay, CNRS/IN2P3, IJCLab, 91405 Orsay, France}
\author{O.~Kamalou}
    \affiliation{Grand Accélérateur National d'Ions Lourds (GANIL), CEA/DRF-CNRS/IN2P3, 
    Bd. Henri Becquerel, 14076 Caen, France}
\author{N.~Kitamura}
    \affiliation{Center for Nuclear Study, University of Tokyo}

\author{V.~Lapoux}
    \affiliation{CEA, Centre de Saclay, IRFU, Service de Physique Nucléaire, 91191 Gif-sur-Yvette, France}
\author{A.~Lemasson}
    \affiliation{Grand Accélérateur National d'Ions Lourds (GANIL), CEA/DRF-CNRS/IN2P3, 
    Bd. Henri Becquerel, 14076 Caen, France}
\author{A.~Matta}
    \affiliation{Department of Physics, Unviversity of Tokyo}
\author{B.~Mauss}
    \affiliation{RIKEN Nishina Center, 2-1, Hirosawa, Wako, Saitama 351-0198, Japan}
    \affiliation{Grand Accélérateur National d'Ions Lourds (GANIL), CEA/DRF-CNRS/IN2P3, 
    Bd. Henri Becquerel, 14076 Caen, France}
\author{P.~Morfouace}
    \affiliation{Grand Accélérateur National d'Ions Lourds (GANIL), CEA/DRF-CNRS/IN2P3, 
    Bd. Henri Becquerel, 14076 Caen, France}
    \affiliation{CEA, DAM, DIF, F-91297 Arpajon, France}
\author{M.~Niikura}
    \affiliation{Department of Physics, Unviversity of Tokyo}
\author{J.~Pancin}
    \affiliation{Grand Accélérateur National d'Ions Lourds (GANIL), CEA/DRF-CNRS/IN2P3, 
    Bd. Henri Becquerel, 14076 Caen, France}
\author{A.~Poves}
    \affiliation{Universidad Autónoma de Madrid,
    MADRID, Spain}
\author{T.~Roger}
    \affiliation{Grand Accélérateur National d'Ions Lourds (GANIL), CEA/DRF-CNRS/IN2P3, 
    Bd. Henri Becquerel, 14076 Caen, France}
\author{T.~Saito}
    \affiliation{National Institute of Advanced Industrial Science and Technology (AIST), Tsukuba 305-8565 - Japan}
\author{C.~Stodel}
    \affiliation{Grand Accélérateur National d'Ions Lourds (GANIL), CEA/DRF-CNRS/IN2P3, 
    Bd. Henri Becquerel, 14076 Caen, France}
\author{J-C.~Thomas}
    \affiliation{Grand Accélérateur National d'Ions Lourds (GANIL), CEA/DRF-CNRS/IN2P3, 
    Bd. Henri Becquerel, 14076 Caen, France}

\date{\today}

\begin{abstract}
   \begin{description}

   \item[Background] A recent sensitivity study has shown that the $^{35}$K$(p,\gamma)^{36}$Ca reaction is 
      one of the ten $(p,\gamma)$ reaction rates that could significantly impact the shape of the calculated X-ray burst light curve. Its reaction rate used up to now in 
      type I X-ray burst calculations was estimated using an old measurement for the mass of $^{36}$Ca and theoretical predictions for the partial decay widths of the first 2$^+$ resonance with arbitrary uncertainties. 
      \item[Purpose] In this work, we propose to reinvestigate the $^{35}$K$(p,\gamma)^{36}$Ca reaction rate, as well as related uncertainties, by determining the energies and decay branching ratios of $^{36}$Ca levels, within the Gamow window of X-ray burst, in the 0.5 to 2 GK temperature range. 
      \item[Method] These properties were studied by means of the one neutron pick-up transfer reaction $^{37}$Ca$(p,d)^{36}$Ca in inverse kinematics using a radioactive beam of $^{37}$Ca at 48 $\mathrm{MeV}$\,nucleon$^{-1}$. The experiment was performed at the GANIL facility using the liquid Hydrogen target CRYPTA, the MUST2 charged particle detector array for the detection of the light charged particles and a zero degree detection system for the outgoing heavy recoil nuclei.  
      \item[Results] The atomic mass of \Ca\ is confirmed and new resonances have been proposed together with their proton decay branching ratios. This spectroscopic information, used in combination with very recent theoretical predictions for the $\gamma$-decay width, were used to calculate the $^{35}$K$(p,\gamma)^{36}$Ca reaction rate. The recommended rate of the present work was obtain within a uncertainty factor of 2 at 1 sigma. This is consistent, with the previous estimate in the X-ray burst temperature range. A large increase of the reaction rate was found at higher temperatures due to two newly discovered resonances.  
      \item[Conclusions] 
      The $^{35}$K$(p,\gamma)^{36}$Ca thermonuclear reaction rate is now well constrained by the present work in a broad range of temperatures covering those relevant to type I X-ray bursts. Our results show that the $^{35}$K$(p,\gamma)^{36}$Ca reaction does not affect the shape of the X-ray burst light curve, and that it can be removed from the list of the few influential proton radiative captures reactions having a strong impact on the light curve.
   \end{description}
\end{abstract}

\maketitle

\section{Introduction}

 Type-I X-ray bursts are among the most energetic events known, which occur in binary systems consisting of a neutron star accreting H/He-rich material from its companion star \cite{Scha06}. 
As the accreted material builds up on the surface of the neutron star, high temperatures and densities (T$_\text{peak}$ $\geq$ 0.9$\times$10$^9$ $\mathrm{K}$ and $\rho$ $\approx$ 10$^6$ g\,cm$^{-3}$) are 
reached. A thermonuclear runaway occurs, leading to a sharp increase of X-ray emission from the star that lasts approximately 
10-100 $\mathrm{s}$.
One of the most important challenges in studying X-ray bursts is understanding the observed luminosity profile, which is 
directly related to the energy released by the nuclear reactions occurring during the thermonuclear explosion. The comparison of 
the observed light curves to the X-ray burst model predictions may be used to constrain the composition of the neutron star's crust as well
as its properties (mass, radius) \cite{Zam12}. X-ray burst models are sensitive to the nuclear reaction rate inputs and recent sensitivity
studies \cite{{Cyb16},{Pari08}} have shown that among the thousands of reactions involved, only the ones participating in the
breakout of the hot-CNO cycle and a few tens of $(\alpha,p)$ and $($p$,\gamma)$ 
reactions have a strong impact on the energy generation of the burst and the final abundances.

The most important $(\alpha,p)$ reactions to be studied are usually 
those involving waiting point nuclei \cite{Cyb16}, where the nuclear reaction flow stalls due to a $(p,\gamma)-(\gamma,p)$ 
equilibrium. This implies to await the $\beta^+$ decay, unless the ($\alpha$,$p$) reaction is fast enough to bypass the waiting point and reach 
higher Z nuclei. The $^{34}$Ar nucleus is such a waiting point and the reaction flow is expected to escape it through an ($\alpha,p)$ reaction, unless a series of two proton captures leading to $^{36}$Ca can compete. 
The $^{35}$K$(p,\gamma$)$^{36}$Ca reaction rate, studied in the present work, has been found to have a significant influence on this reaction pathway, as well as the predicted X-ray burst light curve, when its  nominal value is increased by a factor of one hundred \cite{Cyb16}.

At the typical temperature of an X-ray burst, $T$= 0.5-2~$\mathrm{GK}$ \cite{IlBook}, the Gamow window for the $^{35}$K($p$,$\gamma$)$^{36}$Ca reaction lies between $E_{\mathrm{c.m.}}$=0.37~$\mathrm{MeV}$ and $E_{\mathrm{c.m.}}$=1.93~$\mathrm{MeV}$ (S$_p$=2599.6(61)~$\mathrm{keV}$). It corresponds to excitation energies in the $^{36}$Ca compound nucleus between 2.97 and 4.53 $\mathrm{MeV}$. Considering the $3/2^+$ ground state (g.s.) spin value  of $^{35}$K and the fact that the most relevant proton captures will mostly occur through an \textit{s}-wave ($\ell$ = 0), the resonances of interest in $^{36}$Ca have $J^\pi$=1$^+$,2$^+$.  

Historically, owing to the lack of experimental spectroscopic information on $^{36}$Ca, several studies \cite{Her95,RaTh95,Il01,Wor94}  have estimated the $^{35}$K$(p,\gamma$)$^{36}$Ca reaction rate using theoretical predictions for partial widths and a $2^+$ energy either calculated or adopted from that of the mirror nucleus $^{36}$S  (3.291 MeV). The contribution from the 1$^+$ resonance, identified in the mirror nucleus to be at 4.523 MeV,  was not considered in these works. The reaction $Q$-value was derived from the experimental atomic masses  of $^{36}$Ca and $^{35}$K, known at that time from the $^{40}$Ca($^4$He,$^8$He)$^{36}$Ca \cite{Tr77} and $^{40}$Ca($^{3}$He,$^{8}$Li)$^{35}$K \cite{Ben76} transfer reactions, respectively.

Since then, the excitation energy of the 2$^{+}$ first-excited state in $^{36}$Ca was measured at GANIL \cite{Bu07}, GSI \cite{Do07} and NSCL \cite{Amth08} by  means of  one-neutron knockout reactions from a $^{37}$Ca secondary beam. Taking the most precise measurement, its energy is found to be 3045.0 $\pm$ 2.4 $\mathrm{keV}$ \cite{Amth08}. Moreover, the mass of $^{35}$K was precisely measured using a Penning trap mass spectrometer at ISOLDE \cite{Ya07}. Therefore, the major uncertainty on the reaction $Q$-value, and thereby the $2^+$ resonance energy $E_r^{\mathrm{c.m.}}$ (2$^+)$, came from the uncertainty on the atomic mass of $^{36}$Ca ($\Delta M(^{36}$Ca)=~-6440~$\pm$~40~$\mathrm{keV}$~\cite{Tr77}). The most recent estimate of the reaction rate was made by Iliadis et al. \cite{Il10} by taking into account a 2$^{+}$ state located at 3015 $\pm$ 16 keV~\cite{Do07} corresponding to a resonance energy of $E_r^{\mathrm{c.m.}}$(2$^+)$=459~$\pm$~43~$\mathrm{keV}$.  The presence of this 2$^+$ state at relatively low energy induces an enhanced resonant capture component in the reaction rate, as compared to what was estimated in Fig. 7 of Ref. \cite{Her95}, using $E_r^{\mathrm{c.m.}}$(2$^+)$=700 $\mathrm{keV}$.

While writing the present paper, a more precise mass excess of $\Delta M(^{36}$Ca)= -6483.6~(56) $\mathrm{keV}$ has been obtained  using a Time of Flight - Ion Cyclotron Resonance measurement in a Penning trap \cite{Surb20}.  This leads to a reaction $Q$-value of 2599.6(61) $\mathrm{keV}$ and $E_r^{\mathrm{c.m.}}$(2$^+)$=445 $\pm$ 7 $\mathrm{keV}$, combining the precise mass measurements of  $^{36}$Ca \cite{Surb20} and  $^{35}$K \cite{Ya07}.

In this paper, we report the spectroscopy of $^{36}$Ca through the one neutron pick-up reaction $^{37}$Ca($p,d$)$^{36}$Ca in inverse kinematics. Excitation energies and proton branching ratios of the excited states in or near the Gamow window were obtained. 
Moreover, the mass excess $\Delta M(^{36}$Ca), presently obtained using the reaction $Q$-value of the $($p$,$d$)$ transfer reaction and well-known atomic mass of $^{37}$Ca \cite{Ring07} was compared to the values of Ref.~\cite{Tr77,Surb20}.
These relevant pieces of information are used to better constrain the $^{35}$K($p$,$\gamma$)$^{36}$Ca reaction rate at X-ray burst temperatures and above.

\section{Description of the experiment}
\subsection{Secondary beam production}

The $^{37}$Ca nuclei were produced at GANIL in fragmentation reactions of a 95~MeV$\,$nucleon$^{-1}$ ~$^{40}$Ca$^{20+}$ beam, with an average intensity of $\approx$2~$\mu$Ae,  on a 2 mm $^{9}$Be target. They were separated from other reaction products by the LISE3 spectrometer~\cite{Ann}. A 546~$\mu$m wedge-shaped Be degrader was inserted at the intermediate focal plane to induce a B$\rho$ - $\Delta$E - B$\rho$ selection among the nuclei transmitted after the first magnetic rigidity selection. The Wien filter, located at the end of the spectrometer, was operated at 2400 kV\,m$^{-1}$ in order to induce an additional  velocity selection among the transmitted nuclei. The $^{37}$Ca nuclei were produced at 48 MeV\,nucleon$^{-1}$ with a mean rate of 3500 pps and a purity of 20\%. 

\subsection{Experimental setup}

\begin{figure}[!ht]
  \includegraphics[width=8cm]{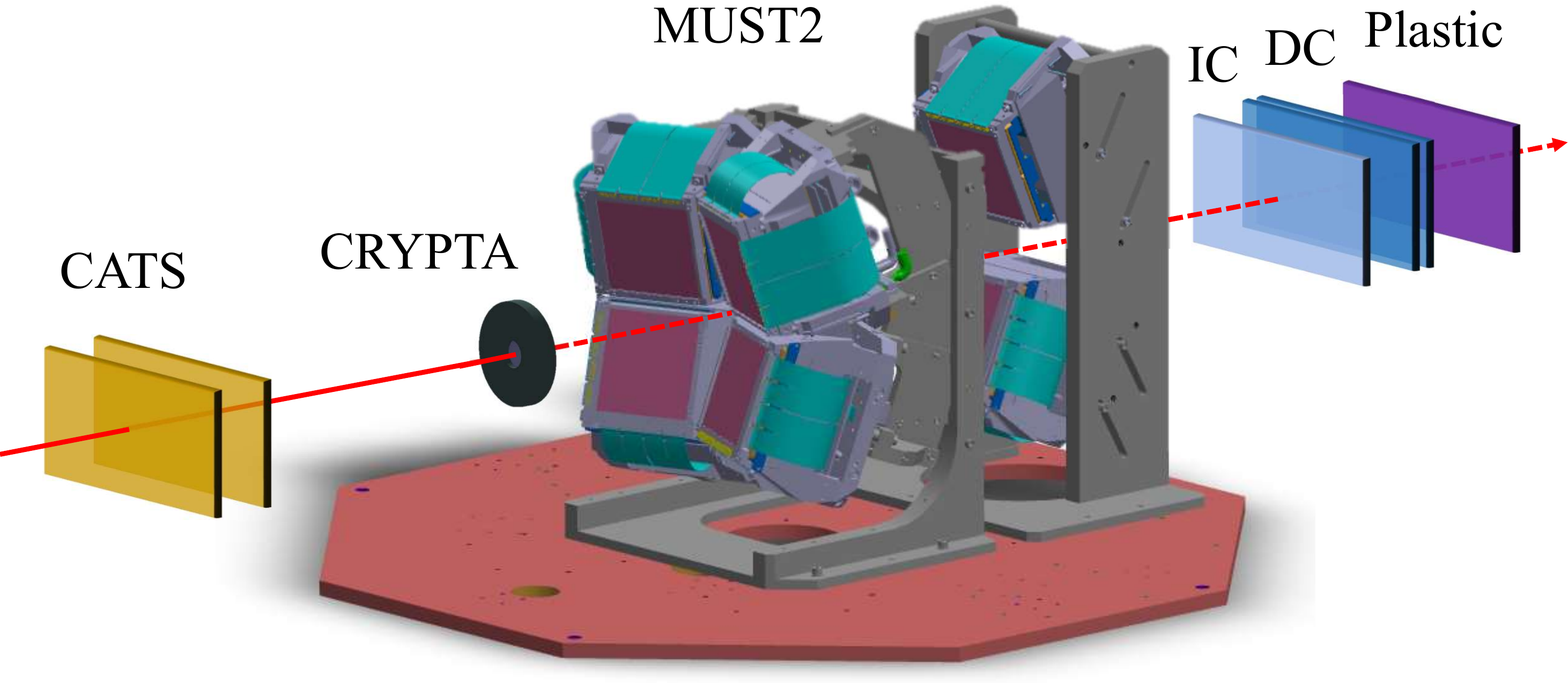}
  \caption{Schematic layout (not on scale) of the experimental setup. The MUST2 telescopes are represented together with the two CATS beam tracker detector, the CRYPTA liquid Hydrogen target, the ionization chamber (IC), the drift chambers (DC), and the plastic scintillator.}
  \label{Setup}
\end{figure}

\begin{figure*}[!ht]
  \includegraphics[height=0.4\textwidth]{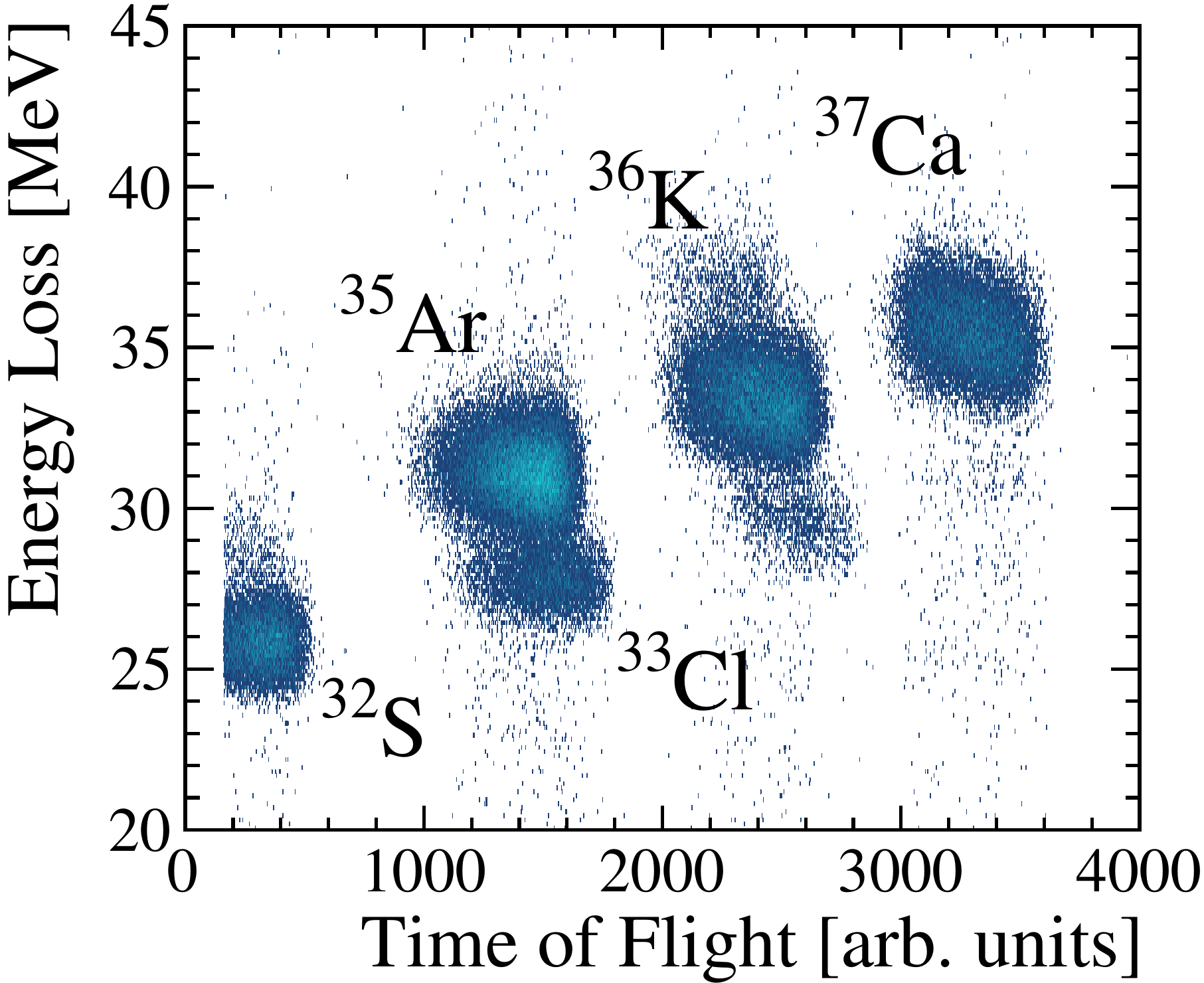}
    \includegraphics[height=0.4\textwidth]{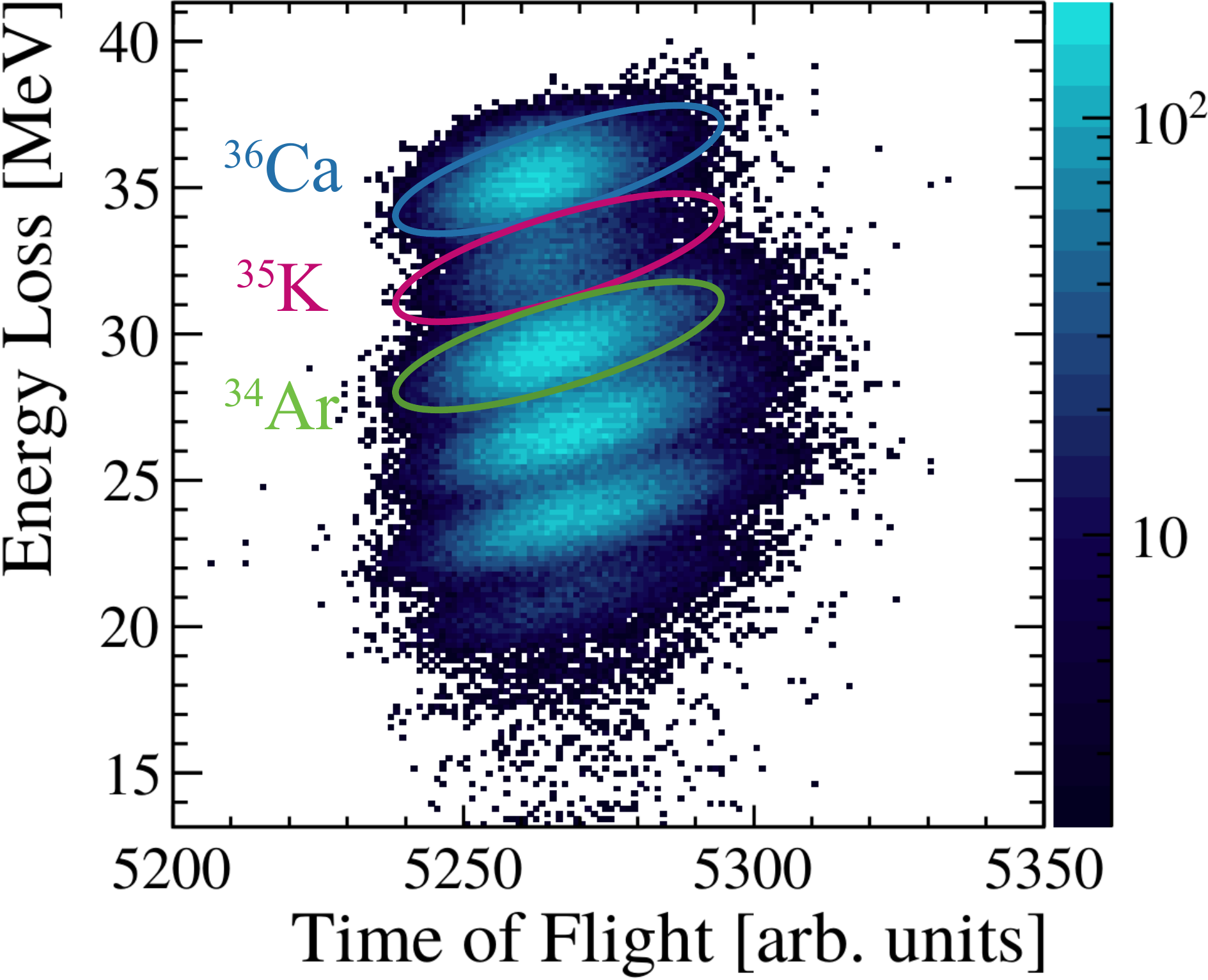}
  \caption{Left: Energy loss vs. TOF identification of the nuclei produced along with $^{37}$Ca. Right: Identification of  heavy transfer residues from their energy loss, measured in the ZDD ionization chamber, and their time-of-flight measured between the first CATS detector and the plastic scintillator located at the end of the ZDD (see text for details). The blue, purple and green contours show the regions corresponding to outgoing Ca, K and Ar nuclei, respectively. }
  \label{ID-ion}
\end{figure*}

Two low-pressure multi-wire devices, CATS \cite{cats}, were placed 67.8 and 118.8 $\mathrm{cm}$ upstream of the liquid Hydrogen target CRYPTA \cite{Koy20}, in order to track the incident ions and determine their interaction point on the target. A schematic layout of the set-up is shown in Fig.~\ref{Setup}. As shown in the left part of Fig. \ref{ID-ion}, the incoming nuclei were identified through their time-of-flight (TOF) measurement, between the CATS detectors and the cyclotron radio-frequency, and their energy loss in an ionization chamber placed at the LISE spectrometer's image focal plane. The main nuclei transmitted along with $^{37}$Ca were $^{36}$K, $^{35}$Ar and $^{32}$S. Another setting of the spectrometer (not shown here) was used to select a cocktail of secondary beam nuclei, shifted by one unit of mass closer to stability, among which the $^{38}$Ca and $^{33}$S nuclei were produced. These nuclei, along with $^{35}$Ar and $^{32}$S, were used to calibrate the detectors with the deuterons emitted in the $($p$,$d$)$ reaction, as described in Sect. \ref{sIII}.  As the $^{37}$Ca nucleus is separated in TOF from the other nuclei, the focal plane ionization chamber was only inserted at the beginning and end of each run to control the incoming beam content.

The liquid Hydrogen ($T$ $\approx$ 18 K) of the CRYPTA target was contained, at a pressure of 0.9 bar, in a 7.6 $\mathrm{cm}$ diameter Al cell with circular apertures of 20 $\mathrm{mm}$ at both ends, closed by 6.47 $\mu$m thick Havar foils. To reach the liquefaction point of the H$_2$ gas (20 K at 1 bar), the vacuum inside the reaction chamber had to be maintained below 2$\times$10$^{-6}$ $\mathrm{mbar}$. Due to the important difference in pressure between the target cell and the reaction chamber, the filling of the target with liquid H$_2$ introduced a significant deformation of the Havar foils. This deformation has been parametrized, using a 10~$\mu$m precision position measurement from a laser system, in order to correct from the energy loss of the particles inside the target event by event. The target thickness spanned from 0.5~mm (at the edges) to 1.5~$\mathrm{mm}$ (at the center), the latter corresponding to an effective thickness of 9.7 mg\,cm$^{-2}$. The target cell was surrounded by a heat shield made of 0.2 $\mu$m aluminized Mylar foils to protect the target cell from radiations. During the experiment, gas contaminants were condensing on the heat shield, forming a layer of ice of unknown thickness and composition (estimated to be less than 10 $\mathrm{\mu}$m equivalent H$_2$O) crossed by the light ejectiles. To minimize this effect and keep the H$_2$ in a liquid phase, the target was warmed-up and cooled down  three times during the 12 days of experiment in order to evaporate the ice layer.

After interaction with the target nuclei, the trajectories of the transfer-like nuclei, their atomic number $Z$ and their time-of-flight (referenced to the CATS detectors) were determined by means of a Zero Degree Detection (ZDD) setup, composed  of an ionization chamber, a set of two XY drift chambers located at 85 $\mathrm{cm}$ and 103 $\mathrm{cm}$ from the target, followed by a 1 $\mathrm{cm}$ thick plastic scintillator. The angular acceptance of the ZDD does not induce kinematical cuts on the detection of the recoil nuclei. Their identification in $Z$ was performed through the measurement of their energy losses in the ionization chamber,  as shown in the right part of Fig. \ref{ID-ion}.

The energy and angle of the light ejectiles were measured by a set of 6 MUST2 telescopes \cite{must2} arranged in the forward direction to cover angles from 3 to 37$^\circ$ in the laboratory frame. Four of them, placed at 22 $\mathrm{cm}$ from the target, were covering angles from 5 to 37$^\circ$ and two more  were placed 57 $\mathrm{cm}$ behind them to cover smaller angles from 3 to 5$^\circ$. For the $^{37}$Ca$($p$,$d$)^{36}$Ca reaction, this corresponds to center-of-mass angles between 2$^\circ$ and 160$^\circ$. 

Each telescope consisted of a 300 $\mu$m thick Double-sided Silicon Stripped Detector (DSSD) with 128 strips on each side, backed by sixteen 4 $\mathrm{cm}$ thick CsI detectors, read out by photodiodes which provide energy-loss ($\Delta E$) and residual energy ($E$) measurements, respectively. 
Light particles identification was obtained from a $\Delta E-E$ matrix for punching through particles. Their total kinetic energy was obtained from the sum of their energy loss in the DSSD  and their residual energy in the CsI crystals, after being corrected by the calculated energy losses in the cryogenic target, its windows and heat shields.
The emission angle of the light ejectiles is deduced from the  information on the impact point of the incident beam on target reconstructed from CATS detector information, and the position measurement of the ejectile in a given strip of the DSSD, with a precision better than 1$^\circ$.

\subsection{Energy Calibrations of the MUST2 detectors} \label{sIII}

Even if the atomic mass of $^{36}$Ca has now been measured with a good accuracy \cite{Surb20}, it is interesting to determine its value with another method, based on transfer reactions.  Even though less precise, this method is more generic and can also be applied to the determination of masses of unbound nuclei. In the present work, the atomic mass and the energy of the excited states of $^{36}$Ca have been determined through the measurement of the energies and angles of the deuterons produced in the $^{37}$Ca$(p,d$)$^{36}$Ca transfer reaction. Moreover, when populating unbound states in $^{36}$Ca, protons are also emitted and their kinematics can be used as well to determine the energy of the resonant states. Thus a precise energy calibration of both deuterons and protons is required in the DSSD as well as in the CsI crystals, in which they are stopped.

The DSSDs were calibrated strip by strip using a mixed alpha source ($^{239}$Pu, $^{241}$Am, $^{244}$Cm) placed at the target position, leading to a global energy resolution of about 40 keV (FWHM) at 5.5 MeV for each telescope.

\begin{figure}[!h]
  \includegraphics[scale=0.5]{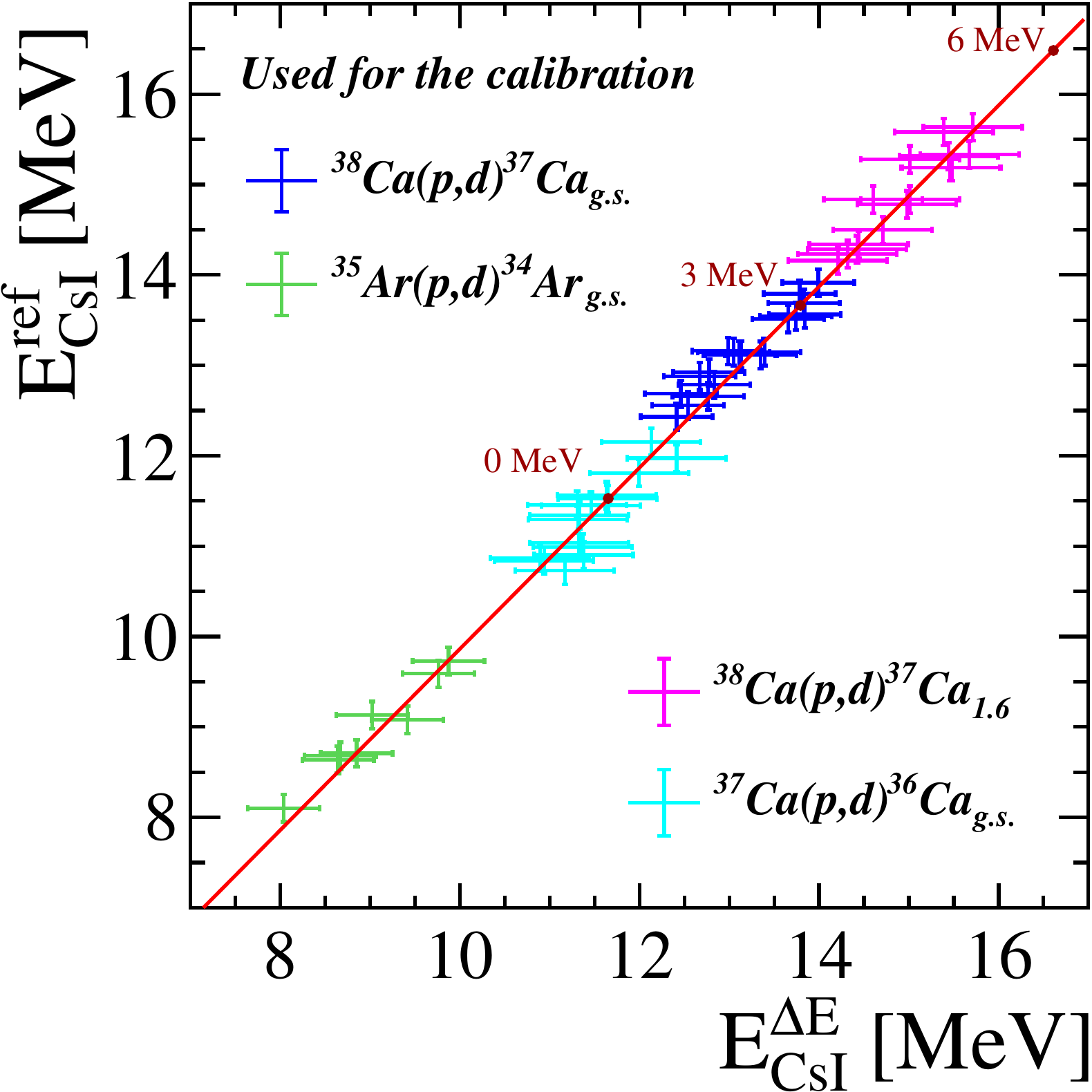}
  \caption{$E^{ref}_{CsI}$  $vs.$ $E^{\Delta E}_{CsI}$ scatter plot for one CsI of one MUST2 telescope. The
  data points from the $^{38}$Ca($p$,$d$)$^{37}$Ca$_{g.s.}$ (blue), $^{35}$Ar($p$,$d$)$^{34}$Ar$_{g.s.}$ (green), $^{37}$Ca($p$,$d$)$^{36}$Ca$_{g.s.}$ (cyan) and $^{38}$Ca($p$,$d$)$^{37}$Ca$_{1.6MeV}$ (magenta) reactions are shown. A linear fit (red curve) was done using the $^{38}$Ca($p$,$d$)$^{37}$Ca$_{g.s.}$ and $^{35}$Ar($p$,$d$)$^{34}$Ar$_{g.s.}$ reactions. The linear fit  describes well the data points from 0 to 6 MeV in excitation energy (written in red) in \Ca ($R^2$=0.992).}
  \label{fig:CalPlot}
\end{figure}

The first step for the energy calibration of the CsI crystals relies on the $E-\Delta E$ correlation of the light particles. The incident energy of each light particle is deduced from its measured energy loss $\Delta E$ in the DSSD. The residual energy in the CsI crystals is subsequently calculated from the measured energy loss in the silicon detector and used to calibrate the CsI crystals. Dead layers are accounted for in this procedure. The correlation between the calculated residual energy (in MeV) and its experimental channel number in the ADC was fitted with a second order polynomial function to determine the calibrated CsI energy $E^{\Delta\text{E}}_{\mathrm{CsI}}$. Given that the CsI crystals show surface and volume inhomogeneities in the light collection (of about $\pm$ 5\%) caused by their aging, each CsI crystal was virtually divided into 64 pixels using the position information from the DSSD.  The energy calibration of the CsI crystals was then done pixel by pixel, eventually covering their full surface. Since the amount of scintillation light produced in a CsI crystal depends on the type of particle, this calibration procedure was performed for the deuterons and the protons independently.

The  second step of the energy calibration of the CsI crystals aimed at correcting the effects of the deformation uncertainties of the target's windows and the non-homogeneity of the dead layers. For this purpose, reference transfer reactions with well known $Q$-values (with less than 2 keV uncertainty) such as $^{38}$Ca$(p,d)^{37}$Ca$_{\text{g.s.}}$ and $^{35}$Ar$(p,d)^{34}$Ar$_{\text{g.s.}}$ were measured during the experiment and used for a refined energy calibration in the energy range of the deuterons resulting from the $^{37}$Ca$(p,d)^{36}$Ca transfer reactions to the ground and excited states up to 6 MeV (see blue, magenta and green crosses in Fig.~\ref{fig:CalPlot}). The reference energy $E^{\text{ref}}_{\text{CsI}}$, calculated event by event, is the expected energy, at the angle where the deuteron was measured, to reconstruct the known mass of the reference nuclei. The error on $E^{\text{ref}}_{\text{CsI}}$ arises from the propagation of the uncertainties on the measured angle. Due to the lack of statistics this second step calibration could only be applied to the CsI crystal and not to each pixel as in the first step. The calibrated values of Fig. \ref{fig:CalPlot} display a linear relationship between  $E^{\text{ref}}_{\text{CsI}}$ and $E^{\Delta \text{E}}_{\text{CsI}}$ over a large range of deuteron energy. Data points corresponding to the population of the $^{36}$Ca g.s. (cyan) are enclosed between the three reference $(p,d)$ reactions. Due to the lack of reference reactions giving rise to a precise determination of the proton energy, the above procedure could only be applied to deuterons. This second step calibration allows to improve the resolution on the excitation energy by 20$\%$ and to reduce the uncertainty on the mass measurement by a factor 3.

\section{Experimental results}

\subsection{Mass excess of $^{36}$Ca} \label{mass36Ca}

The mass excesses of $^{37}$Ca, $^{34}$Ar, and $^{36}$Ca, given in Table \ref{tab:mass}, have been determined from the invariant mass value of their ground state population through $($p$,$d$)$ reactions. The error bars obtained for reference nuclei show the best reachable precision on mass excess measurement with this method, since they are the nuclei used for the calibration.  
The mass excess of \Ca, $\Delta M=-6480(40)$~keV, measured in this work, is in good agreement with the recent measurement, $\Delta M=-6483.6(56)$~keV of Ref.~\cite{Surb20}. As expected, our uncertainty on the $^{36}$Ca mass is larger than the one obtained from the penning trap measurement \cite{Surb20}, but similar to that obtained in another transfer reaction  \cite{Tr77}. This uncertainty is dominated by systematic errors arising from multiple effects such as the propagation of errors on the measured angle and energy of the deuteron and on the energy calibration of the CsI. They have been estimated combining the standard deviation of  independent measurements performed using the 4 MUST2 telescopes, located at the closest distance from the target. Taking the most precise atomic mass values of $^{36}$Ca and $^{35}$K, the proton separation energy of $^{36}$Ca is deduced to be $S_p$ = 2599.6(61)~keV. 

\begin{table}[!h]
\caption{Mass excesses ($\Delta M$), obtained in the present work for $^{37}$Ca, $^{34}$Ar, and $^{36}$Ca using the $($p$,$d$)$ reaction are compared to other experimental works. As derived from $Q$-values, our  results use the precise experimental atomic masses of $^{38}$Ca \cite{Ring07}, $^{35}$Ar \cite{mass}, and $^{37}$Ca\cite{Ring07}, respectively.}

\centering
\small\addtolength{\tabcolsep}{+7pt}
\setlength\extrarowheight{2pt}
\begin{tabular}{ccc}
\hline
\hline
Nucleus & $\Delta M$ (keV)  & $\Delta M$ (keV) \\
& this work& literature  \\  \hline
$^{36}$Ca   & -6480(40)         & \begin{tabular}[c]{@{}l@{}}-6450(40)\cite{Tr77};\\ -6483.6(56)\cite{Surb20}\end{tabular}             \\ 
$^{37}$Ca   &  -13141(13)        & -13136.1(6) \cite{Ring07}         \\
$^{34}$Ar   &   -18403(25)       &  -18378.29(8) \cite{Herf01}          \\ \hline
\hline
\end{tabular}
\label{tab:mass}
\end{table}

\subsection{Excited states in $^{36}$Ca}\label{Ex} \label{E36Ca}

\begin{figure}[!h]
\includegraphics[width=8.5cm]{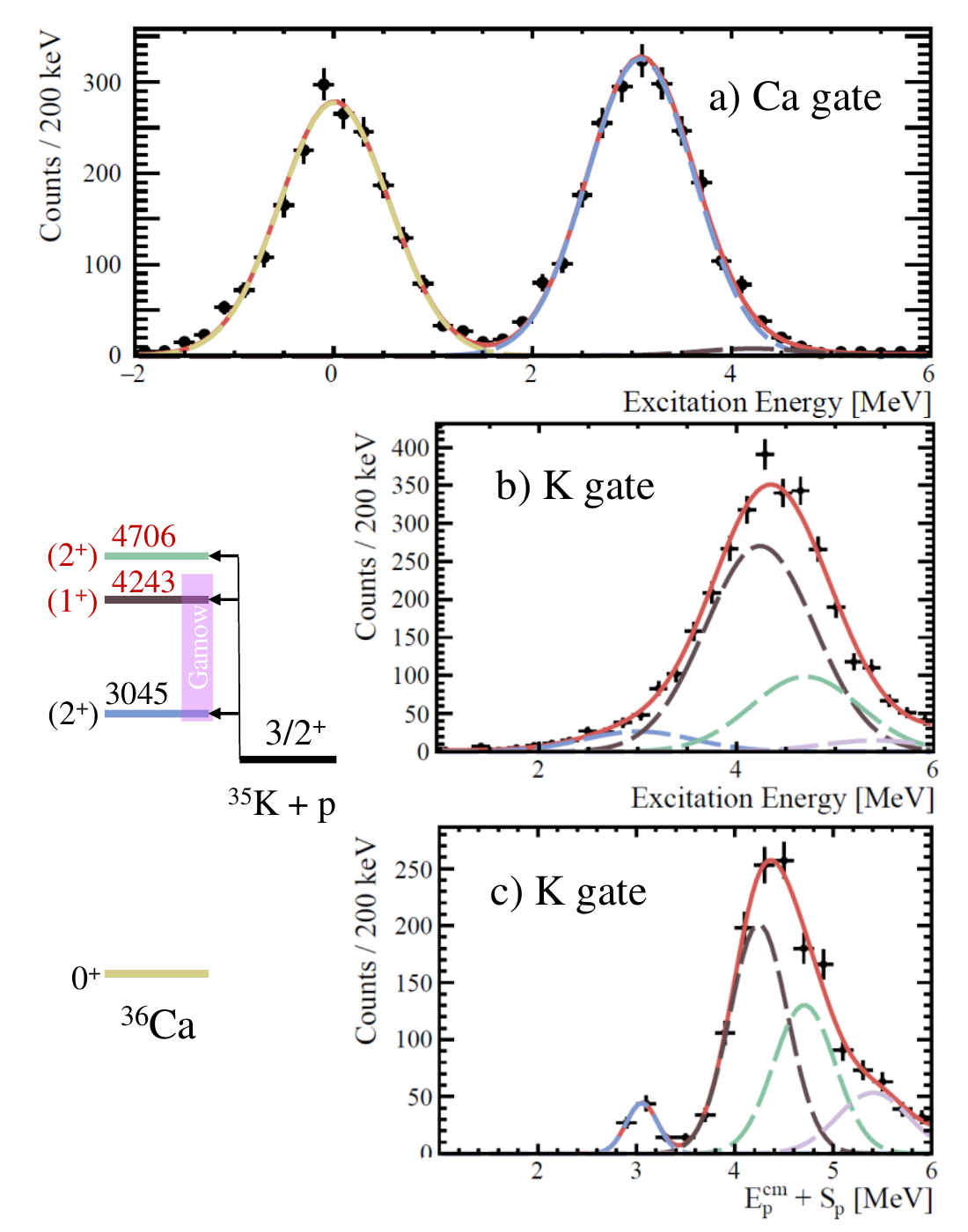}

    \caption{Excitation energy spectrum of \Ca reconstructed from the measurement of energy and angle of the deuteron and gated on outgoing Ca a) or K b) nuclei in the ZDD system. The center-of-mass energy of the protons emitted from unbound states in \Ca has been added to the one-proton separation energy in the excitation energy spectrum $E_p^{\mathrm{c.m.}}$ of c). In all spectra, the red lines display the best fits (p-value = 0.67 for $E_x$ and 0.82 for $E_p^{\mathrm{c.m.}}$) and colored dashed-lines represent the different resonances in \Ca used for the fit. Those especially relevant  for X-ray bursts are shown in the level scheme on the left with the same color codes. Energy of exited states are taken from Ref.~\cite{Amth08} for the 2$^+_1$ state and from our work otherwise (highlighted in red).}
  \label{fig:Ex}
\end{figure}

 The missing mass method has been applied in inverse kinematics to determine the excitation energy ($E_x$) of the states produced in $^{36}$Ca. After gating on an incoming $^{37}$Ca the excitation energy is reconstructed from  the total kinetic energy and the angle of the deuterons produced in the $(p,d)$ reaction. Figures \ref{fig:Ex}a) and \ref{fig:Ex}b) display the $E_x$ spectra gated on the outgoing Ca or K nuclei in the ZDD (as shown in the right part of  Fig.\ref{ID-ion}), respectively. The  fit of the excitation energy spectrum has been performed using multiple Gaussian function, assuming that the natural width of the states is much smaller than the experimental resolution. The red lines in Figures \ref{fig:Ex}a) and \ref{fig:Ex}b) show the best total fits obtained and the colored dashed lines are the individual state contributions used for each fit. All the parameters of the fit are free except the resolution. The energy-dependence of the peak widths was estimated using the $nptool$ package~\cite{Matta2016}, in which the whole set-up was simulated. The resolution was then strongly constrained in the fit, using the reference width of the known and isolated ground state and the simulated energy-dependence. The number of contributions used in the fit was guided by the number of levels populated in the mirror reaction \cite{Gray70} and by the statistical test of the p-value.
 
  The peaks corresponding to the feeding of the ground  and  first 2$^{+}$ excited states in $^{36}$Ca are well identified in Fig.~\ref{fig:Ex}a). As expected, the peak corresponding to the g.s. disappears when gated on K nuclei. The energy of the 2$^{+}$ state is measured at 3059 (30) keV in Fig.~\ref{fig:Ex}a) and 2982 (120) keV in Fig. \ref{fig:Ex}b) (blue curve), in agreement with the value of 3045 (2.4) keV \cite{Amth08}, within the error bars. The relatively large uncertainties arise from a nearby resonance, as will be discussed below.  As the  2$^{+}$ state is unbound with respect to one and two proton emissions, a certain fraction of its decay occurs to the ground state of $^{35}$K, bound by only 83.6 (5) keV \cite{mass}, with the emission of a proton. This is  discussed in the following.
 
In Fig. \ref{fig:Ex}c), the one-proton energy spectrum $E_p^{\mathrm{c.m.}}$ has been reconstructed in the $^{36}$Ca  center-of-mass from the measured  energy and angle of the proton in coincidence with the deuteron of the $(p,d)$ reaction and the outgoing K nuclei. For convenience, the one-proton separation energy ($ S_{p}$(\Ca ) = 2599.6(61)~keV) has been added in Fig. \ref{fig:Ex}c) to the proton energies to ease the comparison with  the excitation energy spectra of Figs. \ref{fig:Ex}a,b). The resulting excitation energy  resolution is 2 to 4 times better when reconstructed with the protons than with the deuterons: 130 keV at $E_x$= 3 MeV and 300 keV at 5 MeV with the protons and an almost constant value around 550 keV with the deuterons. This effect arises from the more forward focused kinematics of the protons, as compared to deuterons. In addition, as the proton energy spectrum is constructed with less than half of the CsI crystals,  the systematic uncertainty caused by their inhomogeneities is smaller in the $E_p^{\mathrm{c.m.}}$ spectrum. 

Thus, the peak corresponding to the 2$^+_1$ state is better separated from the others in the $E_p^{\mathrm{c.m.}}$ spectrum  of Fig. \ref{fig:Ex}c), as compared to the excitation energy peak shown in Fig. \ref{fig:Ex}b). Note also that the triple coincidence (deuteron, proton and K nucleus) cleans the $E_p^{\mathrm{c.m.}}$ spectrum from all type of background. The fit of the $E_p^{\mathrm{c.m.}}$ spectrum has been performed using multiple Gaussian functions, whose energy-dependent widths have been constrained from simulations, assuming again that their natural width is much smaller than the experimental resolution. The energy of the 2$^+_1$ state is found at 3057 (20) keV. Its uncertainty comes from the moderate statistics. The presently determined  2$^{+}_1$ energy agrees well with the ones of 3036(11) \cite{Bu07} and 3045(2.4)\cite{Amth08} keV, determined by $\gamma$-decay, as well as the value of 3059 (30) keV derived from our fit of the excitation energy spectrum of Fig.~\ref{fig:Ex}a).

As shown in Fig. \ref{fig:Ex}c), several states are identified up to 6 MeV. One of them at $E_{x} = 4.243(40)$ MeV lies inside the Gamow window of the $^{35}$K$(p,\gamma$)$^{36}$Ca reaction. This state is also visible at a compatible energy  in the excitation energy spectrum of Fig. \ref{fig:Ex}b). According to the selection rules of the $(p,d)$ transfer reaction, $J^\pi$=1$^{+}$ and 2$^{+}$ states are populated with a $(s_{1/2})^{1}(d_{3/2})^{1}$  configuration in the removal of an $\ell$=0 neutron from the $2s_{1/2}$ orbital. This assumption is confirmed by the fact that one $J^\pi$=1$^{+}$ and two 2$^{+}$ states are populated in the same energy range in $^{36}$S by means of the mirror reaction $^{37}$Cl($d$,$^{3}$He)$^{36}$S \cite{Gray70}. The isobaric analogue 1$^{+}_1$ state was measured at $E_x(1^+)$= 4523.0 (6) keV. Therefore we tentatively assign a spin-parity of 1$^{+}$ for the excited state of \Ca at 4.243(40) MeV. Two shell model calculation was performed, one in the $sd$ valence space with USDB interaction and the other using  the  full $sdpf$ valence space with sdpfu-mix plus Coulomb interaction. Calculation in $sd$ valence space predicts the position of this 1$^{+}_1$ state in \Ca at $E_{x} = 4161$ keV and while calculations in $sdpf$ valence space predicts $E_{x} = 4000$ keV.

Given that the energy resolution of the proton spectrum is accurately determined from simulations, two states are needed between 4 and 5 MeV to achieve the best fit of the spectrum. Besides the 1$^+$ state at 4.24~MeV discussed above, a (2${^+_2}$) state has been identified at E$_{x}$ = 4.706 (100) MeV in \Ca, close to the upper border of the Gamow window. The identification of two states (tentatively 1${^+}$ and 2${^+_2}$) in this energy range is reinforced by a similar observation in the mirror nucleus $^{36}$S, with a (1${^+}$, 2${^+_2}$) doublet at 4.523 and 4.572 MeV, fed within similar relative intensities in the  $^{37}$Cl($d$,$^{3}$He)$^{36}$S reaction. The energy and feeding of these states, obtained from Fig. \ref{fig:Ex}c), are compatible with those obtained independently in the excitation energy spectrum of Fig. \ref{fig:Ex}b) from the detection of deuterons.

Other states are observed between 5 and 8 MeV, but since they are well outside of  the Gamow window, they will have no impact on the reaction rate and will not be discussed here. As a last remark, despite the fact that all states discussed here (2$^{+}_1$,1$^{+}$,2$^{+}_2$)  are unbound by two-proton emission ($S_{2p} \approx$ 2683 keV), no peak is observed at the corresponding resonance energy in the Ar-gated $E_x$ spectrum (not shown here). Therefore, we shall neglect their $2p$ decay contribution to determine their proton to $\gamma$ decay branches  in the following section.

\begin{table*}[]
\caption{The experimental and theoretical results for the resonant states in \Ca obtained in this work are presented. Tentative spins and parities $J^{\pi}$, measured excitation energies $E_x$ (in keV) and proton branching ratios are listed for the three states identified in \Ca. Results of Shell Model calculations for partial $\gamma$-width (meV), proton spectroscopic factors $C^2S$ and proton width (in meV) as well as their corresponding proton branching ratios are presented. Two different shell model calculations have been performed, one restricted to the $sd$ orbitals and USDB interaction, the other using the full $sdpf$ valence space with sdpfu-mix plus Coulomb interaction. The results are compared to the shell model results of Ref. \cite{Her95}. Predicted widths, obtained for a given calculated excitation energy, have been renormalized to the experimental values given in the second column. The proton spectroscopic factors are given for the orbital which contributes the most to the proton width ($i.e.$ $s_{1/2}$ orbital for all states)}
\centering
\small\addtolength{\tabcolsep}{+1pt}
\setlength\extrarowheight{3pt}
\begin{tabular}{cccccccccccccccccc}
\hline
\hline
\multicolumn{13}{c}{Present work}                                                             &                      & \multicolumn{4}{c}{Herndl \cite{Her95}} \\ 
\cline{1-13}
\cline{15-18}
\multicolumn{3}{c}{Exp.}                               & & \multicolumn{4}{c}{$sd$ shell} && \multicolumn{4}{c}{$sdpf-mix$ shells} && \multicolumn{4}{c}{$sd$}     \\
\cline{1-3}
\cline{5-8}
\cline{10-13}
\cline{15-18}
J$^{\pi}$       & E$_x$     & $B_p$   && $\Gamma_{\gamma}$ & $C^2S_{1/2}$ & $\Gamma_p$     &$B_p$       &&  $\Gamma_{\gamma}$  & $C^2S_{1/2}$& $\Gamma_p$      &$B_p$       && $\Gamma_{\gamma}$ & $C^2S_{1/2}$& $\Gamma_p$  & $B_p$      \\ \hline
(2$^{+}_{1}$) & 3045(2.4) &0.165(10)&& 0.5               &0.009         &  0.87          & 0.64       && 0.99\footnotemark[1]& 0.009       & 0.84            & 0.46       && 0.4               & 0.009       & 0.94        & 0.70    \\
(1$^{+}$)     & 4243(40)  & $>$0.96 &&  37.1             &0.0009        & 2.8$\times$10$^4$& $\approx 1$&&  65.4               & 0.002       & 6.3$\times$10$^4$ &$\approx$ 1 &&                   &             &             &        \\
(2$^{+}_{2}$) & 4706(100) & $>$0.97 && 0.2               &              &                &            && 7.4                 &  0.003      & 3.3$\times$10$^5$ &$\approx$ 1 &&                   &             &             &        \\ \hline
\hline
\end{tabular}
\footnotetext[1]{from \cite{VD18}}
\label{tab:BR}
\end{table*}

\subsection{Experimental proton branching ratios} \label{BR}

The first 2$^{+}$ excited state of \Ca at $E_x=3.045$~MeV has been identified both in the excitation energy spectra gated by Ca and K, meaning that it decays through $\gamma$-ray  and proton emission with respective partial widths $\Gamma_\gamma$ and $\Gamma_p$. We propose here to determine its experimental proton branching ratio B$_p$=$\Gamma_p$/$\Gamma_{tot}$, with $\Gamma_{tot}$=$\Gamma_\gamma$+$\Gamma_p$, using two methods.  As for the $1^+ $ and $2^+_2$ states, no counts are observed in the $\gamma$ decay branch, we shall determine minimum values of B$_p$, based on the fit of the K-gated and Ca-gated excitation energy spectra.

\begin{figure}
    \centering

        \includegraphics[width=0.45\textwidth]{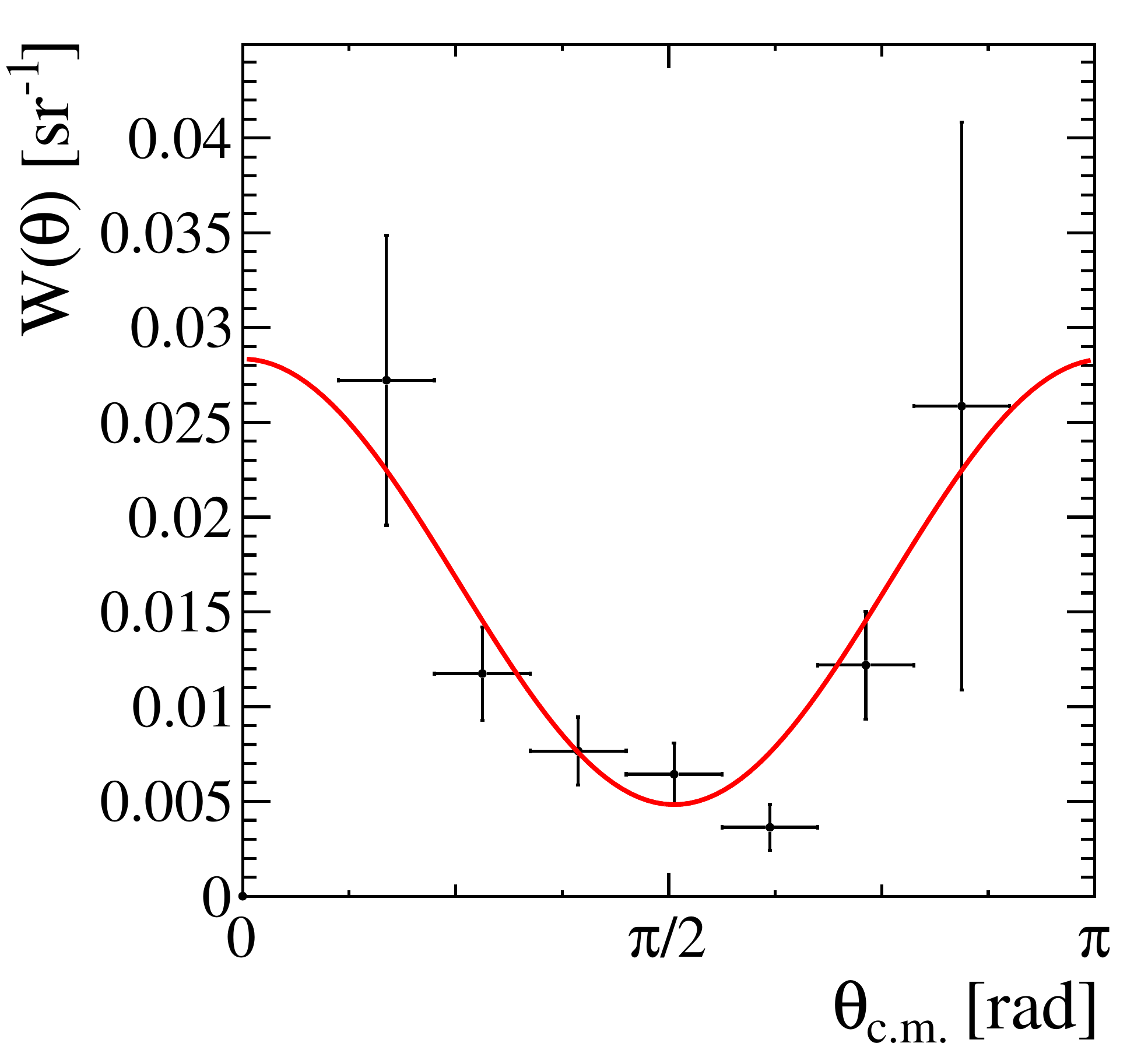}
 
    \caption{Proton-deuteron angular correlation corresponding to the 2$^+_1$ state as a function of the center-of-mass angle of the proton in the \Ca frame. The red line shows the best fit obtained with $k_{max}$=2 (p-value = 0.52).} 
  \label{fig:AD}
\end{figure}

\noindent {\it First method-} The experimental proton-deuteron angular correlation is shown in Fig.~\ref{fig:AD} as a function of the proton center-of-mass emission angle in the \Ca\ frame. This correlation function is obtained after normalisation of the number of decaying protons per angle by the total number of counts observed in the excitation energy peak and correction of the relative geometrical efficiency between protons and deuterons (which have significantly different angular distributions). The geometrical efficiency was computed using the {\it nptool} simulation package where the $^{37}$Ca($p$,$d$)$^{36}$Ca transfer reaction and the subsequent proton emission were both simulated with an isotropic angular distribution. It has been restricted to events corresponding to proton center-of-mass energies ranging from 2.5 to 3.5 MeV to focus on the study of the 2$^+_1$ decay. Errors for each point of the angular correlation are dominated by statistical uncertainties.

This correlation function $W(\theta)$ can be described by a sum of even Legendre polynomials, $P_k(cos(\theta))$~\cite{Pronko}:
\begin{equation}
    \label{eq:corr}
     W(\theta) = \sum_{k=0}^{k_{max}} A_kP_k(cos(\theta)),
\end{equation}
where $A_k$ are coefficients obtained from a fit to the experimental angular correlation. The sum is truncated at a maximum value of $k_{max} = min(\ell+\ell',2J)$, where $\ell$ and $\ell'$ are the possible proton orbital angular momenta, and $J$ is the spin of the decaying state. The value of $k_{max}$=2, which results from the best fit shown in Fig.~\ref{fig:AD}, can then be used to constrain the spin assignment of the decaying \Ca\ state. Given the fact that the ground state of $^{35}$K has $J^\pi=3/2^+$, this implies that the first excited state in \Ca\ has either  $J=1$ or $J=2$ assignment. This is in agreement with the $J^\pi=2^+$ value expected from the mirror nucleus and shell model calculations. 

By integrating the angular correlation function  over the full $4\pi$ solid angle, a proton branching ratio of $B_p = \Gamma_p/\Gamma_{tot}=0.16$~(2) is determined. The uncertainty results from the error propagation of the fitted parameters.

\noindent {\it Second method-} By counting the number of events associated to the feeding of the first 2$^{+}$ excited state in the K-gated and Ca-gated excitation energy spectra, a $\Gamma_p/\Gamma_\gamma=0.21$~(3) ratio is found. Using the fact that: 

\begin{equation} \label{Bp}
B_p = 1 / (1 + \Gamma_\gamma/\Gamma_p), 
\end{equation} 

\noindent a branching ratio of $B_p = 0.17$~(2) is obtained. The uncertainty is due to the fitting of the peaks in the excitation energy spectrum, as well as in the choice of the gates in Ca and K in the ZDD spectrum of Fig. \ref{ID-ion}. The proton branching ratio values for the 2$^+$ state derived with the two methods are in excellent agreement. 

The second method was applied to compute the following $B_p$ lower limits: $>$ 0.96 for the 1$^{+}$ state and $>$ 0.97 for the 2$^{+}_{2}$ at one $\sigma$ confidence level. The fit of the Ca-gated excitation energy spectrum of Fig. \ref{fig:Ex}a) was used to estimate their maximum contributions to the $\gamma$-decay channel, such as the small one to the 1$^{+}$ state, displayed in brown color.

As shown in Table \ref{tab:BR}, the weighted average value  $B_p=0.165$~(10) for the 2$^+$ state is found to be significantly weaker than all theoretical predictions, while the deduced $B_p$ lower limits for the higher 1$^{+}$ and 2$^{+}_{2}$ states are consistent with the predictions.

\subsection{Calculated gamma widths} \label{GW}

The predicted $\Gamma_{\gamma}$ partial widths of Table \ref{tab:BR}, expressed in meV, were computed using the following relation \cite{IlBook}:
\begin{equation}
    \label{eq:gwidth}
    \Gamma_\gamma(\omega L) = \frac{8\pi(L+1)}{L[(2L+1)!!]^2} \left(\frac{E_{\gamma}}{\hbar c}\right)^{2L+1}B(\omega L),
\end{equation}

\noindent where $\omega$ names the nature of the transition (Electric or Magnetic) and $L$ its multipolarity. $B(\omega L)$ is the reduced transition probability for the $\gamma$ decay of the resonant state (in unit of $e^2\,fm^4$ for E2 transitions and $\mu_N^2$ for M1 transitions) and $E_\gamma$ the energy of the $\gamma$-ray transition. Shell model calculations usually predict $B(\omega L)$ and $E_\gamma$ values, from which $\Gamma_{\gamma}$ is calculated. However, as the experimental excitation energy of the 2$^+_1$ state is known and differs from the calculated ones, the predicted partial widths $\Gamma_{\gamma}$ listed in Table \ref{tab:BR} are obtained from Eq. \ref{eq:gwidth} using experimental energies and calculated $B(E2)$ values.

Two different shell model calculations have been performed in the present work, one  restricted to the $sd$ orbitals and USDB interaction, the other using the full $sdpf$ valence space with sdpfu-mix plus Coulomb interaction. The reduced transition probabilities, which contribute the most to the $\gamma$-ray width of each state, obtained in $sd$ valence space are: $B(E2;2^+_1\rightarrow0^+_{g.s.})=2.4~\mathrm{e^2\,fm^4}$, $B(M1;1^+\rightarrow0^+_{g.s.})=0.01~\mathrm{\mu_N^2}$, $B(M1;1^+\rightarrow2^+_{1})=1.4~\mathrm{\mu_N^2}$, $B(M1;2^+_2\rightarrow2^+_1)=0.002~\mathrm{\mu_N^2}$, $B(E2;2^+_2\rightarrow0^+_{g.s.})=0.02~\mathrm{e^2\,fm^4}$. Values obtained in $sdpf$ shell are  $B(E2;2^+_1\rightarrow0^+_{g.s.})=4.7~\mathrm{e^2\,fm^4}$, $B(M1;1^+\rightarrow0^+_{g.s.})=0.04~\mathrm{\mu_N^2}$, $B(M1;1^+\rightarrow2^+_{1})=1.5~\mathrm{\mu_N^2}$, $B(M1;2^+_2\rightarrow2^+_1)=0.06~\mathrm{\mu_N^2}$, $B(E2;2^+_2\rightarrow0^+_{g.s.})=2.2~\mathrm{e^2\,fm^4}$.

The major difference between the shell model calculations presented here or in Ref.~\cite{Her95}, resides in the size of their valence spaces: when restricted to $sd$ shells the proton core is closed, while the use of a broader $sdpf$ valence space allows proton excitations. When using the $sd$ valence space, the 2$^+_1$ state in $^{36}$Ca ($^{36}$S) is of pure neutron (proton) origin. It follows that the $B(E2)$ values of the two mirror nuclei can simply be derived from their squared neutron to proton effective charges ratio, $B(E2)(^{36}$Ca)=  e$^2_n$/e$^2_p$ B(E2)($^{36}$S), where e$_n $ (e$_p$) are the neutron (proton) effectives charges usually adopted to be 0.5 (1.5). 

As the  2$^+_1$ state in $^{36}$Ca is certainly not totally of pure neutron origin, the calculated  $\Gamma_{\gamma}$ using a $sd$ valence space ($\approx$ 0.4 meV) represents a lower limit. At the other extreme, a maximum $\Gamma_{\gamma}$ of about 3.7 meV is obtained for $^{36}$Ca when assuming the same $B(E2)$ value as in the mirror nucleus $^{36}$S, after correcting from their different 2$^+_1$ energies. This latter assumption would imply that the 2$^+$ state has a very mixed (and similar) structure in both nuclei. This is very unlikely for two reasons. First, the two nuclei are likely doubly magic, at least based on the high excitation energy of their first excited states. Second, the 2$^+_1$ state in $^{36}$S is very well populated by the $^{37}$Cl($d$,$^3$He)$^{36}$S proton removal reaction, with spectroscopic factors values for the 2$^+_1$ and 1$^+_1$ states (0.86 and 0.75 respectively \cite{Gray70}) that are close to the single particle values, meaning it has a strong proton component rather than a mixed proton and neutron one.

\subsection{Calculated proton widths} \label{PW}

The proton widths $\Gamma_p$ of the states listed in Table \ref{tab:BR} are obtained by multiplying their single-particle width $\Gamma_{sp}$ with the spectroscopic factor $C^2S$: 
\begin{equation}
    \label{eq:pwidth}
\Gamma_p = \Gamma_{sp} \times C^2S.
\end{equation}

The $C^2S$ values are obtained from shell model calculations, while $\Gamma_{sp}$ are calculated by scattering phase shifts in a Woods-Saxon potential \cite{WSP} whose depths are adjusted to match the experimental resonance energies. The Wood-Saxon potential parameters used for calculation can be found in page 239 of Ref. \cite{BM98}. In the present work, the widths of the 2$^+_1$ state obtained in the $sd$ and $sdpf$ shell model calculations agree very well with each other, while those for the 1$^+$ state differ by more than a factor two.

It is important to note that the $\Gamma_{p}$ values are obtained by multiplying a very small $C^2S$ number (of the order of 10$^{-3}$) by large  barrier penetrability factors for the protons, especially for those having $\ell>0$. Despite this, the $\Gamma_{p}$ values obtained with the two calculations agree reasonably well. The $C^2S$ values are small as they correspond to the emission of a proton from an almost pure $1p1h$ neutron state, selectively populated here by the $($p$,$d$)$ transfer reaction.

\section{The $^{35}$K($p$,$\gamma$)$^{36}$Ca reaction rate}

The thermonuclear reaction rate per particle pair 
is given by \cite{LONGLAND20101}: 
\begin{equation}
    \label{eq:rate}
 <\sigma \nu > = {\left(\frac{8}{\pi\mu}\right)}^{1/2} \frac{1}{(kT)^{3/2}} \int_{0}^{\infty} E\sigma(E)e^{-E/kT}dE,
\end{equation}
where $\mu$ is the reduced mass of the two interacting nuclei, 
$k$ is the Maxwell-Boltzmann constant, $T$ is the temperature in Kelvin, $E$ is the center-of-mass energy in MeV and $\sigma(E)$ is the nuclear reaction cross section in barn.

The $^{35}$K$(p,\gamma)^{36}$Ca reaction rate depends on resonant capture (RC) and direct capture (DC) contributions, that we shall discuss in the following.

\subsection{Resonant capture}

In the presence of  narrow resonances, the reaction rate can be expressed as: 
\begin{equation}
    \label{eq:narrow}
    <\sigma \nu > = \frac{1.5399\times10^{11}}{N_{A}} \left(\frac{\mu}{T_{9}}\right)^{3/2}   \sum_{i} (\omega\gamma)_{i}e^{-11.605E_{i}/T_{9}},
\end{equation}

where 

\begin{equation}
 \label{eq:og}
(\omega\gamma)_{i}= \frac{2J_{i}+1}{(2J_p+1)(2J_{^{35}K}+1)}\frac{\Gamma_{\gamma,i}\Gamma_{p,i}}{\Gamma_{i}},
\end{equation}
is the resonance strength of the $i^{th}$ resonance with $\Gamma_{\gamma,i}$, $\Gamma_{p,i}$ and $\Gamma_{i}$ its partial $\gamma$-ray,  proton and total width in MeV, respectively, $E_{i}$ the resonance energy in MeV, $J_{i}$ the spin of the resonance, $J_p$ and $J_{^{35}K}$ are the proton spin (1/2) and the g.s. spin of $^{35}$K (3/2), respectively. $T_9$ is the temperature in GK and $\mu$ is the reduced mass. This assumption of narrow resonance is valid as the resonant states considered here have a total width far below their resonance energies.

As shown in Eq.~\ref{eq:narrow}, the resonance energy, the spin, as well as the total and partial widths of all resonances are needed to compute the reaction rate. The resonance energy $E_{r}$ for the 2$^+_1$ state has been determined from the excitation energy of Ref. \cite{Amth08} (being the most precise measurement performed by $\gamma$-ray spectroscopy) and the recent mass measurement of Ref. \cite{Surb20}. For the 1$^+$ and 2$^+_2$ states, excitation energies are the one determined in the present work. The spin values used for the computation are the ones proposed in Sect. \ref{Ex}.   

As we could only determine precisely the proton branching ratio in the present work (and only a lower limit for the 1$^+$ and 2$^+_2$ states), we choose to fix the $\Gamma_\gamma$ partial widths using the $sdpf$ shell model calculation  which makes use of the broadest valence space and correctly reproduces the energy of the first 2$^+$ state. Once $\Gamma_\gamma$ is fixed, $\Gamma_p$ and $\Gamma_{tot}$ can be derived for the 2$^+_1$ state using the experimental $B_p$ value as the proton and $\gamma$ decays are the only open channels. 

As for the 1$^{+}$ and the 2$^{+}_{2}$ resonances, the proton partial width dominates the total width. It follows that  the resonance strength of the Eq.~\ref{eq:og} can be approximated by $\omega\gamma \simeq \frac{\Gamma_{\gamma}}{8}(2J_r+1)$, with $J_r$ the spin of the resonance. All the resonance parameters needed to compute the reaction rate are listed in Table.~\ref{tab:Inputs_exp} 

The reaction rate has been computed using the Monte-Carlo code RatesMC \cite{LONGLAND20101}, allowing a statistically meaningful evaluation of the reaction rate based on experimental and/or theoretical uncertainties. A Gaussian probability density function is assumed for the resonance energies and a log-normal distribution is used as a probability density function for $\gamma$-width and proton-width values. 

A central value $\Gamma_{\gamma}$ of 0.99 meV was used for the 2$^+_1$ state with an uncertainty factor of 1.7, which corresponds to values between 0.58 and 1.7 meV at one sigma. This way, we accept the lower (0.4 meV) and upper limit (3.7 meV) of $\Gamma_{\gamma}$, discussed in Sect. \ref{GW}, at about 2$\sigma$. The same uncertainty factor is assumed for the $\Gamma_{\gamma}$ widths of the 1$^+$ and the 2$^+_2$ states. The uncertainty on $\Gamma_p$ of the 2$^+_1$ is deduced from that on $\Gamma_{\gamma}$ and on the experimental $B_p$ value, following Eq. \ref{Bp}.

\begin{table}[!h]
 \caption{Resonances parameters used in this work to compute the $^{35}$K($p$,$\gamma$)$^{36}$Ca reaction rates. Resonance spin-parity value, experimental energy and calculated $\gamma$-width (using the $sdpf$ valence space) are given in the three first columns. The proton-width $\Gamma_p$, derived from the calculated $\gamma$-width and the experimental proton branching ratio for the first 2$^+$ state, is given in the fourth column. The resonant strength of each state is listed in column five.}

\centering
\small\addtolength{\tabcolsep}{+5pt}
\setlength\extrarowheight{5pt}
\begin{tabular}{ccccc}
\hline
\hline
$J^{\pi}$   & $E_{r}$ [keV]  & $\Gamma_{\gamma}$ [meV]   &  $\Gamma_p$ [meV] &$\omega \gamma$ [meV]\\ 
\hline
(2$^{+}$)     & 445 (7)        &  0.99\footnotemark[1]     &0.20 &0.102(50) \\
(1$^{+}$)     & 1643 (41)      &  65.4\footnotemark[1]     & & 25 (14)  \\
(2$^{+}_{2}$) & 2106 (100)     &  7.4\footnotemark[1]     &&4.6 (25) \\
\hline
\hline
\end{tabular}
\footnotetext[1]{with a uncertainty factor 1.7} 
\label{tab:Inputs_exp}
\end{table}

\begin{figure}[!htb]
    \centering
\includegraphics[width=0.45\textwidth]{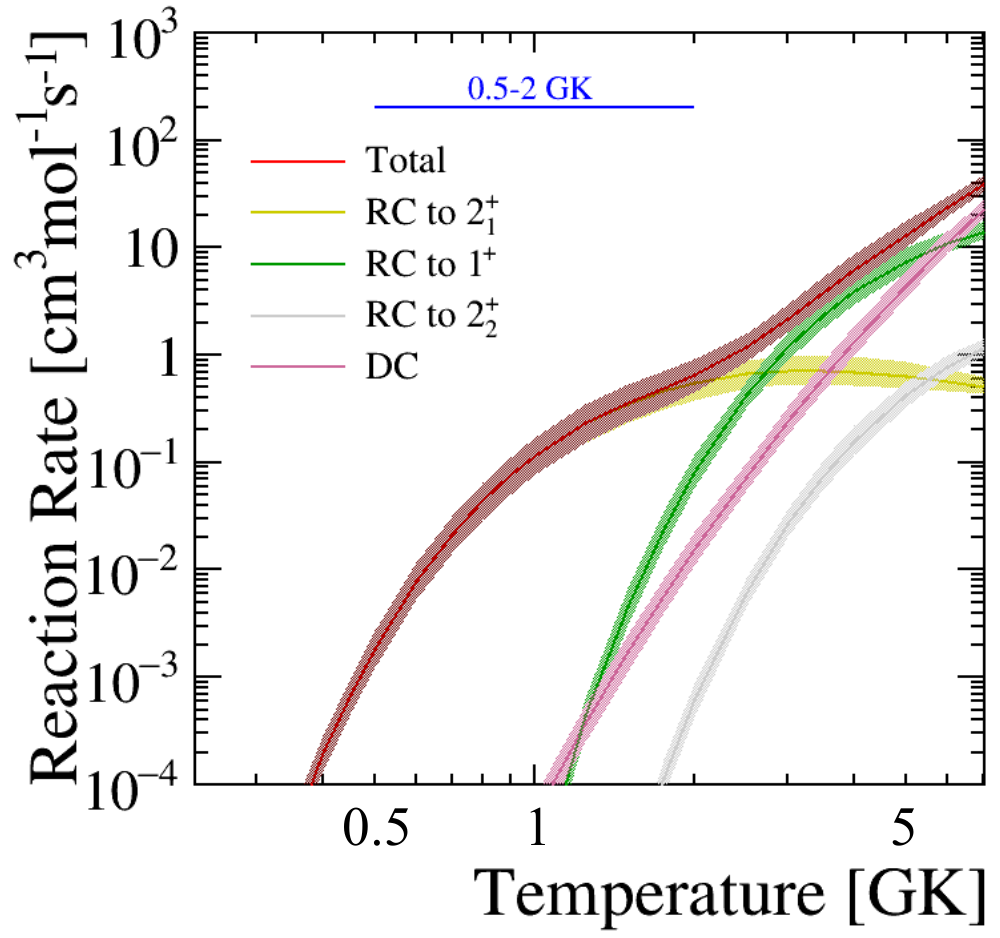} \\
\includegraphics[width=0.45\textwidth]{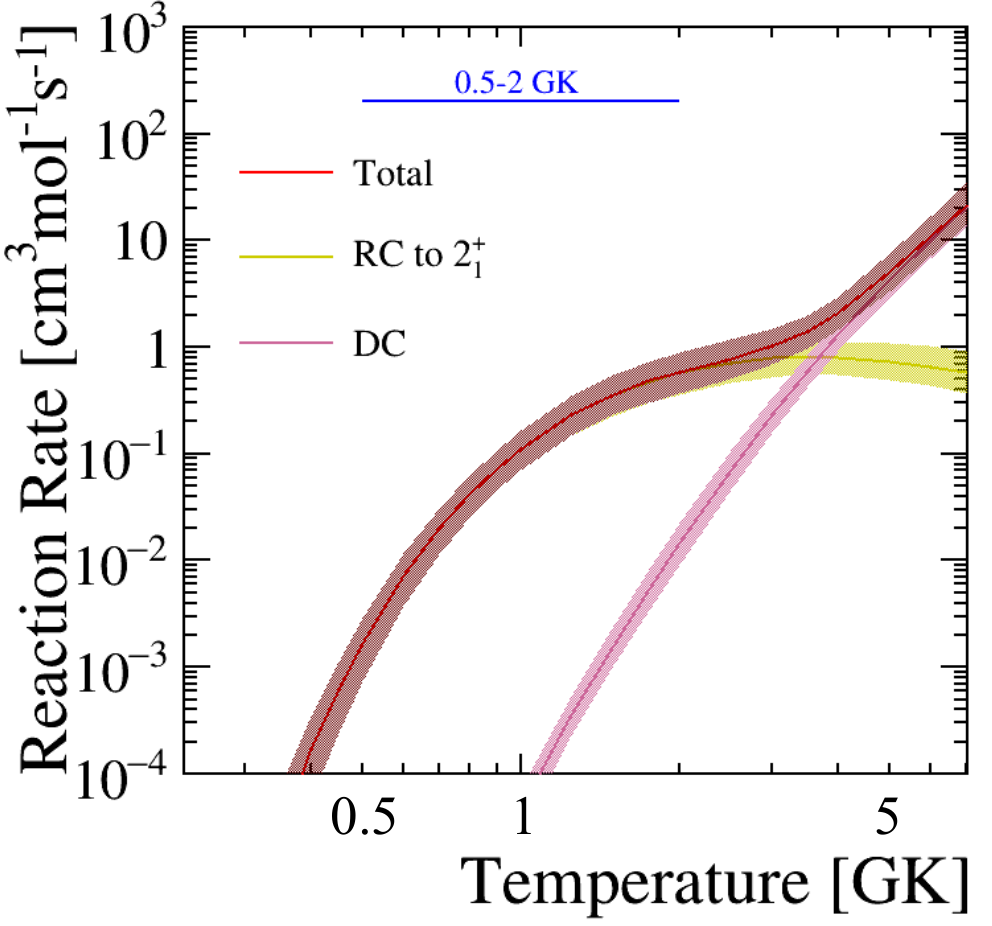}   
\caption{ The $^{35}$K($p$,$\gamma$)$^{36}$Ca reaction rate  calculated (in units of cm$^{3}$\,mol$^{-1}$\,s$^{-1}$) in this work (top) and  in \cite{Il10} (bottom). The red curve represent the total reaction rate, which includes the contributions of resonant (RC) and direct (DC) captures. The thickness of the curves represents a coverage probability of 68$\%$. The indicated range of 0.5 - 2 GK is typical of X-ray bursts temperatures.} 
  \label{fig:RR}
\end{figure}

\subsection{Direct capture}

The DC component mostly occurs through an $\ell$=2 proton capture between the ground state of  $^{35}$K and that of $^{36}$Ca, which have $J^\pi$= 3/2$^+$ and 0$^+$, respectively. In a simplified picture, 3 protons are occupying the $1d_{3/2}$ orbital in the g.s. of $^{35}$K and one more can occupy it by the direct capture process to the ground state of \Ca. The calculations of Ref. \cite{Her95} predict a proton spectroscopic factor $C^2S_p$ for \Ca of 3.649, while the $sdpf$ shell model calculation predicts a slightly smaller value of 3.37. We propose the use of the mean value between the two (3.5) and an uncertainty of 0.15 to agree, within one sigma, with the two calculations.

\subsection {Discussion}

The evolution of the calculated rates with temperature for the $^{35}$K($p$,$\gamma$)$^{36}$Ca reaction is shown in Fig \ref{fig:RR}. The top part displays the results obtained when using the presently discussed resonances and corresponding widths, while the bottom part shows the results obtained from the work of Iliadis et al. \cite{Il10}, in which only the $2^+$ resonance was considered (with the following parameters: $E_{r}=459(43)$ keV, $\Gamma_{\gamma} = 0.38$~meV and $\Gamma_p = 1.4$~meV) in addition to the DC part. In the X-ray burst temperature range (0.5-2 GK), the reaction rate is dominated by this resonance. Therefore the major differences between our work and the one of Ref. \cite{Il10} reside in the values used for the partial $\gamma$-width (to be discussed in the next paragraph), the partial proton-width, the resonance energy of the 2$^+$ state and their associated uncertainties.  

For the partial $\Gamma_{\gamma}$ width, an arbitrary value of 50\% was assumed in \cite{Il10} with a central value of 0.4 meV for the 2$^+$ state. It corresponds to a log-normal confidence interval for the widths of [0.25-0.64] meV at one sigma, which partly overlaps with our current determination of the $\gamma$-ray partial width. The uncertainty on the 2$^+$ resonance energy considered in the present work was reduced using the very recent precise measurement of the mass of \Ca from \cite{Surb20} and the excitation energy from \cite{Amth08}. As shown in Fig. \ref{fig:RR}, the contributions of the 1$^+$ and 2$^+_2$ resonances  to the total reaction rate, not taken into account in \cite{Il10}, start to be significant at temperatures above $T=2$~GK.

\begin{figure}
  \includegraphics[width=0.45\textwidth]{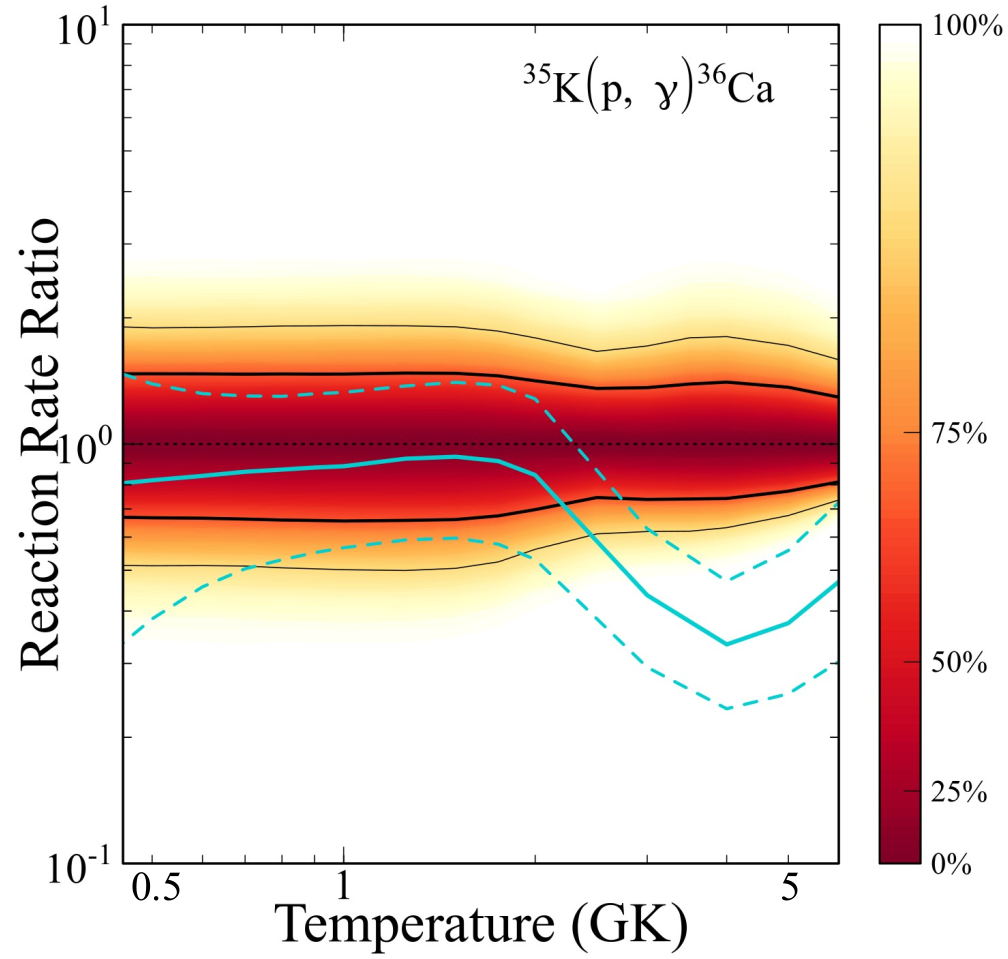}
  \centering
  \caption{Ratio of rates normalized to our recommended reaction rate. The area delimited by the thick/thin black lines and centered around 1 represent the recommended value at the 68\% and 95\% confidence levels, respectively. The thick and dashed blue lines correspond to the reaction rate given by Iliadis et al. \cite{Il10} and at the 68\% confidence level, normalized to our recommended value. 
  }
  \label{fig:Ratio}
\end{figure}

The ratio of the calculated reaction rate by Iliadis et al. \cite{Il10} to our recommended value (given numerically in Tab. \ref{tab:RRNum}) is shown in Fig. \ref{fig:Ratio}. The colored areas outlined by the thick/thin black lines show the uncertainty on the recommended reaction rate calculated in this work with a coverage probability of 68\% and 95\% respectively. The thick and dashed blue lines correspond to the reaction rate given in \cite{Il10} with the associated 68\% uncertainties respectively, normalized to our recommended reaction rate.  For the temperature range of interest, the results are similar. We have also estimated that the contributions to the 0$^+_2$ ($\ell=2$ proton capture) and 3$^-$ ($\ell=1$) states, not identified here but present in the mirror nucleus, are negligible in the Gamow window. At temperatures higher than 2 GK, our recommended reaction rate is systematically higher due to the contributions of the 1$^+$ and 2$^+_2$ resonances, not included in \cite{Il10}. This reaction rate should, however, be considered as a lower limit, as higher-energy resonances may additionally contribute to the reaction rate beyond 2 GK.

\begin{table}[]
\small\addtolength{\tabcolsep}{+5pt}
\setlength\extrarowheight{1.1pt}
  \caption{ Low, recommended and high thermonuclear rates
of the $^{35}$K($p$,$\gamma$)$^{36}$Ca reaction (in units of cm$^{3}$\,mol$^{-1}$\,s$^{-1}$) as a function of temperature. Interval between low and high rates represents a confidence level of 68\% (1 $\sigma$). }

\begin{tabular}{cccc}
\hline
\hline
T [GK]                   & Low                  & Recommended          & High \\ \hline

 0.010 &  3.191$\times10^{-51}$ &          3.350$\times10^{-51}$ &        3.516$\times10^{-51}$ \\   
 0.011 &  2.345$\times10^{-49}$ &          2.461$\times10^{-49}$ &        2.591$\times10^{-49}$ \\   
 0.012 &  1.051$\times10^{-47}$ &          1.104$\times10^{-47}$ &        1.160$\times10^{-47}$ \\   
 0.013 &  3.151$\times10^{-46}$ &          3.316$\times10^{-46}$ &        3.487$\times10^{-46}$ \\   
 0.014 &  6.785$\times10^{-45}$ &          7.123$\times10^{-45}$ &        7.487$\times10^{-45}$ \\   
 0.015 &  1.100$\times10^{-43}$ &          1.157$\times10^{-43}$ &        1.218$\times10^{-43}$ \\   
 0.016 &  1.407$\times10^{-42}$ &          1.478$\times10^{-42}$ &        1.555$\times10^{-42}$ \\   
 0.018 &  1.281$\times10^{-40}$ &          1.346$\times10^{-40}$ &        1.414$\times10^{-40}$ \\   
 0.020 &  6.238$\times10^{-39}$ &          6.547$\times10^{-39}$ &        6.886$\times10^{-39}$ \\   
 0.025 &  1.498$\times10^{-35}$ &          1.573$\times10^{-35}$ &        1.654$\times10^{-35}$ \\   
 0.030 &  5.656$\times10^{-33}$ &          5.950$\times10^{-33}$ &        6.257$\times10^{-33}$ \\   
 0.040 &  3.229$\times10^{-29}$ &          3.393$\times10^{-29}$ &        3.565$\times10^{-29}$ \\   
 0.050 &  1.507$\times10^{-26}$ &          1.585$\times10^{-26}$ &        1.665$\times10^{-26}$ \\   
 0.060 &  1.633$\times10^{-24}$ &          1.719$\times10^{-24}$ &        1.808$\times10^{-24}$ \\   
 0.070 &  6.881$\times10^{-23}$ &          7.243$\times10^{-23}$ &        7.612$\times10^{-23}$ \\   
 0.080 &  1.503$\times10^{-21}$ &          1.581$\times10^{-21}$ &        1.663$\times10^{-21}$ \\   
 0.090 &  2.050$\times10^{-20}$ &          2.152$\times10^{-20}$ &        2.262$\times10^{-20}$ \\   
 0.100 &  2.112$\times10^{-19}$ &          2.274$\times10^{-19}$ &        2.501$\times10^{-19}$ \\   
 0.110 &  2.632$\times10^{-18}$ &          3.680$\times10^{-18}$ &        5.509$\times10^{-18}$ \\   
 0.120 &  6.334$\times10^{-17}$ &          1.066$\times10^{-16}$ &        1.794$\times10^{-16}$ \\   
 0.130 &  1.416$\times10^{-15}$ &          2.425$\times10^{-15}$ &        4.077$\times10^{-15}$ \\   
 0.140 &  2.174$\times10^{-14}$ &          3.653$\times10^{-14}$ &        6.068$\times10^{-14}$ \\   
 0.150 &  2.322$\times10^{-13}$ &          3.847$\times10^{-13}$ &        6.272$\times10^{-13}$ \\   
 0.160 &  1.837$\times10^{-12}$ &          3.007$\times10^{-12}$ &        4.807$\times10^{-12}$ \\   
 0.180 &  5.706$\times10^{-11}$ &          9.077$\times10^{-11}$ &        1.426$\times10^{-10}$ \\   
 0.200 &  8.717$\times10^{-10}$ &          1.370$\times10^{-09}$ &        2.113$\times10^{-09}$ \\   
 0.250 &  1.116$\times10^{-07}$ &          1.720$\times10^{-07}$ &        2.577$\times10^{-07}$ \\   
 0.300 &  2.683$\times10^{-06}$ &          4.073$\times10^{-06}$ &        6.069$\times10^{-06}$ \\   
 0.350 &  2.502$\times10^{-05}$ &          3.786$\times10^{-05}$ &        5.595$\times10^{-05}$ \\   
 0.400 &  1.298$\times10^{-04}$ &          1.955$\times10^{-04}$ &        2.894$\times10^{-04}$ \\   
 0.450 &  4.570$\times10^{-04}$ &          6.891$\times10^{-04}$ &        1.018$\times10^{-03}$ \\   
 0.500 &  1.227$\times10^{-03}$ &          1.856$\times10^{-03}$ &        2.738$\times10^{-03}$ \\   
 0.600 &  5.214$\times10^{-03}$ &          7.901$\times10^{-03}$ &        1.163$\times10^{-02}$ \\   
 0.700 &  1.413$\times10^{-02}$ &          2.145$\times10^{-02}$ &        3.173$\times10^{-02}$ \\   
 0.800 &  2.911$\times10^{-02}$ &          4.418$\times10^{-02}$ &        6.535$\times10^{-02}$ \\   
 0.900 &  4.983$\times10^{-02}$ &          7.581$\times10^{-02}$ &        1.125$\times10^{-01}$ \\   
 1.000 &  7.534$\times10^{-02}$ &          1.150$\times10^{-01}$ &        1.712$\times10^{-01}$ \\   
 1.250 &  1.518$\times10^{-01}$ &          2.313$\times10^{-01}$ &        3.445$\times10^{-01}$ \\   
 1.500 &  2.356$\times10^{-01}$ &          3.564$\times10^{-01}$ &        5.288$\times10^{-01}$ \\   
 1.750 &  3.295$\times10^{-01}$ &          4.886$\times10^{-01}$ &        7.116$\times10^{-01}$ \\   
 2.000 &  4.551$\times10^{-01}$ &          6.538$\times10^{-01}$ &        9.253$\times10^{-01}$ \\   
 2.500 &  8.745$\times10^{-01}$ &          1.197$\times10^{+00}$ &        1.653$\times10^{+00}$ \\   
 3.000 &  1.618$\times10^{+00}$ &          2.196$\times10^{+00}$ &        3.118$\times10^{+00}$ \\   
 3.500 &  2.790$\times10^{+00}$ &          3.810$\times10^{+00}$ &        5.535$\times10^{+00}$ \\   
 4.000 &  4.507$\times10^{+00}$ &          6.124$\times10^{+00}$ &        8.876$\times10^{+00}$ \\   
 5.000 &  9.892$\times10^{+00}$ &          1.291$\times10^{+01}$ &        1.811$\times10^{+01}$ \\   
 6.000 &  1.872$\times10^{+01}$ &          2.316$\times10^{+01}$ &        3.048$\times10^{+01}$ \\   
 7.000 &  3.204$\times10^{+01}$ &          3.773$\times10^{+01}$ &        4.696$\times10^{+01}$ \\   
 8.000 &  5.114$\times10^{+01}$ &          5.785$\times10^{+01}$ &        6.893$\times10^{+01}$ \\   
 9.000 &  7.719$\times10^{+01}$ &          8.471$\times10^{+01}$ &        9.723$\times10^{+01}$ \\   
10.000 &  1.104$\times10^{+02}$ &          1.196$\times10^{+02}$ &        1.330$\times10^{+02}$ \\

\hline
\hline
\end{tabular}
 \label{tab:RRNum}

\end{table}

The sensitivity study of Cyburt et al. \cite{Cyb16} concluded that the $^{35}$K($p$,$\gamma$)$^{36}$Ca reaction would lead to a significant modification of the X-ray burst light curve if the reaction rate was a factor of one hundred higher than that of Iliadis et al. \cite{Il10}. Such an increase is absolutely ruled out by our study for which a factor of 3 difference is found at maximum for the reaction rate between 0.5 and 2 GK. 

\section{Conclusion}

The spectroscopy of \Ca was investigated via the one neutron pick-up reaction $^{37}$Ca$(p,d)^{36}$Ca in inverse kinematics, in view of determining useful information for the $^{35}$K$(p$,$\gamma$)$^{36}$Ca reaction rate and compare it to earlier works such as \cite{Il10}. The \Ca atomic mass was measured and matches very well with previous values \cite{Tr77,Surb20}. The energy of the first 2$^+$ excited state was confirmed and new resonances have been reported in the vicinity of the Gamow window, at excitation energies $E_x = 4.243(40)$ and $4.706(100)$ MeV. Based on shell model calculations in the $sdpf$ valence space and the comparison to the mirror nucleus ($^{36}$S), spins  and parities 1$^+$ and 2$^+$ were proposed for these two new observed states, respectively. The proton branching ratio $B_p = 0.165(10)$ of the first 2$^+$ state was measured with two independent methods and lower  limits, $B_p(1^+) > 0.96$ and $B_p(2^+_2) > 0.97$ were estimated for the two other resonant states.

A Monte Carlo procedure \cite{LONGLAND20101}, which consistently takes into account the uncertainties on the energy, spin parity, partial and total widths of the \Ca states, was then used to calculate the $^{35}$K($p$,$\gamma$)$^{36}$Ca reaction rate between 0.01 and 10 GK with its corresponding uncertainty. Shell model predictions of $B(\omega L)$ were used to compute the only non experimentally-constrained parameter for the resonant states: $\Gamma_\gamma$. The factor 1.7 uncertainty associated to this prediction dominates the total uncertainty of the reaction rate in the X-ray burst temperature range of interest. Therefore, the determination of the gamma width (or lifetime) of the 2$_1^+$ state is still desirable, as it would provide an almost fully experimentally-constrained reaction rate.

The present work provides constrains in a broad range of temperatures for the $^{35}$K($p$,$\gamma$)$^{36}$Ca reaction rate. It should be noted, however, that some additional contributions (not studied in this work) may further increase the reaction rate above 2 GK. Up to 4~GK, our recommended value is consistent, within one sigma,  with the one  of \cite{Il10}, previously used in X-ray burst models. Based on the reaction sensitivity tests of  Cyburt et al. \cite{Cyb16}, our measured reaction rate is not sufficiently different from previous estimation to modify significantly the luminosity profile of X-ray burst. Therefore, the $^{35}$K($p$,$\gamma$)$^{36}$Ca reaction can be removed from the list of the proton radiative captures reactions having a strong impact on the light curve.


\begin{acknowledgments}
The continued support of the staff of the GANIL facility is gratefully acknowledged. We thank Richard Longland and Phil Adsley for their help with the use of the RatesMC code. We acknowledge significant support from NFS grant PHY-1811855. 
\end{acknowledgments}

\newpage
\bibliographystyle{apsrev4-1}
\bibliography{36CaAstro_final}

\end{document}